\documentclass[12 pt,aps,prd,showpacs]{revtex4-1}   
\usepackage{natbib}
\usepackage{amsmath}    
\usepackage{graphicx}   
\usepackage{epsfig}
\usepackage{epsfig}
\usepackage{graphicx}
\usepackage[latin1]{inputenc}
\usepackage{float}
\usepackage{color}
\usepackage{amsmath,amssymb}
\usepackage{mathtools}
\usepackage{natbib}

\def\b{\begin{equation}}
\def\e{\end{equation}}

\newcommand{\defeq}{\vcentcolon=} 
\begin{document}
\title{New signatures of the dynamical Casimir effect in a superconducting circuit}
\author{Andreson L. C. Rego}
\email{andreson@if.ufrj.br}
\affiliation{Instituto de F\'\i sica, Universidade Federal do Rio de Janeiro, 21945-970, Rio de Janeiro, Brazil}
\author{Hector O. Silva}
\email{hosilva@phy.olemiss.edu}
\affiliation{Department of Physics and Astronomy, The University of Mississippi, University, MS 38677, USA}
\author{Danilo T. Alves}
\email{danilo@ufpa.br}
\affiliation{Faculdade de F\'\i sica, Universidade Federal do Par\'a, 66075-110, Bel\'em, Brazil}
\author{C. Farina}
\email{farina@if.ufrj.br}
\affiliation{Instituto de F\'\i sica, Universidade Federal do Rio de Janeiro, 21945-970, Rio de Janeiro, Brazil}

\date{\today}
\begin{abstract}
We found new signatures of the dynamical Casimir effect (DCE) in the context of superconducting circuits. We show that if the recent experiment made by Wilson {\it et al}, which brought the DCE into reality for the first time, is repeated with slight modifications
(for instance, different values for the capacitance of the SQUID),
three remarkable results will show up, namely:
{\it (i)} a quite different spectral distribution for the created particles, deviating from the typical parabolic shape;
{\it (ii)} an enhancement by a factor of approximately $5 \times 10^3 $  in the number of created particles with half driven frequency of the effective moving mirror and
{\it (iii)} an enhancement by a factor of $3 \times 10^2$ in the particle creation rate. These results may guide the experimentalists in their search for alternative routes to observe the DCE in future experiments.
\end{abstract}
%
\maketitle

The theoretical prediction of the dynamical Casimir effect (DCE) - the creation of real particles from the vacuum of quantized fields induced by its interaction with moving bodies - was made by Moore \cite{Moore-1970} approximately four and a half decades ago.
This phenomenon was also investigated by DeWitt \cite{Dewitt-PhysRep-1975} and Fulling and Davies \cite{Fulling-Davies-PRS-1976-I-Fulling-Davies-PRS-1977-I}.
One main difficulty in observing such an effect
experimentally, where the effort in putting material mirrors into motion is limited to
non-relativistic velocities, was already realized by this author, as it is evident in his own words: ``the creation of photons from zero-point energy is altogether negligible''. Although there are some experimental proposals for the detection of the DCE involving real mechanical motion of boundaries (see for instance \cite{Kim-Brownell-Onofrio-2006}), several authors followed an idea first proposed by Yablonovitch \cite{Yablonovitch-1989} which consists in \textit{simulating} fast-moving mirrors by a physical mechanism which gives rise to a time-dependent boundary condition (BC) imposed on the quantum field by a \textit{static} mirror \cite{experimentos-Braggio,J-R-Johansson-G-Johansson-C-Wilson-F-Nori-PRL-2009-PRA-2010,Dezael-Lambrecht-EPL-2010,Kawakubo-Yamamoto-PRA-2011,Faccio-Carusotto-EPL-2011,C-Wilson-et-al-Nature-2011, Lahteenmaki-et-al-PNAS-2013}.
One of these  proposals, based on a superconducting coplanar waveguide terminated at a SQUID (Superconducting Quantum Interference Device),
which simulates a single moving mirror whose effective velocity may achieve   $\approx 10\%$ of the speed of light in vacuum,
led to the report by Wilson {\it et al} of the first observation of the DCE \cite{C-Wilson-et-al-Nature-2011}.
In this experiment, a time-dependent magnetic flux is applied to the SQUID, changing
its effective inductance and resulting in a time-dependent BC. As a consequence, the coplanar waveguide
becomes equi\-va\-lent to a one-dimensional transmission line with variable length.
A second observation of the DCE was reported by L\"{a}hteenm\"{a}ki {\it et al} \cite{Lahteenmaki-et-al-PNAS-2013}. These authors observed the DCE by means of periodical changes in the index of refraction of a microwave cavity (idealized by two transmission lines separated by a capacitive gap with an array of SQUIDs placed at one of the extremities of these strip-lines) with a Josephson metamaterial.

In the experiment described in \cite{C-Wilson-et-al-Nature-2011}, Wilson and collaborators observed, for a harmonically oscillating Josephson energy
with angular frequency $\omega_0$, a spectral distribution for the created photons with an approximate parabolic shape with a maximum value at $\omega_0/2$,
as predicted theoretically by these authors (see Fig. 2 of the first paper and Table III of the second reference in \cite{J-R-Johansson-G-Johansson-C-Wilson-F-Nori-PRL-2009-PRA-2010}). Due to the particular values for the experimental parameters, which allowed certain approximations to be made, this prediction is in agreement with an earlier paper by Lambrecht {\it et al} \cite{Lambrecht-Jaekel-Reynaud-PRL-1996}.
%
Calculations of spectra for a moving cavity can be found in \cite{Lambrecht-et-al-EPJD-1998}.
The detection of such a parabolic spectral distribution, with the apparatus cooled enough to allow an unambiguous distinction between
dynamical Casimir photons and thermal photons, was one of the main goals in the experiment described in \cite{C-Wilson-et-al-Nature-2011}.

In the present letter, our purpose is to show that if some of the parameters used in the recent experiment made by Wilson {\it et al} \cite{C-Wilson-et-al-Nature-2011} (for instance, different values for the capacitance of the SQUID) are
appropriately adjusted, three remarkable results will show up, namely:
{\it (i)} a breakdown of the usual parabolic spectral distribution into a two peaked one;
 {\it (ii)} an enhancement by a factor of $\approx 5 \times 10^3$ in the number of created particles with half driven frequency of the effective moving mirror and
{\it (iii)} an enhancement by a factor of $\approx3\times 10^2$ in the particle creation rate. Furthermore, we also provide results for the number of created particles per unit frequency, around half the driven frequency ($\omega_0/2$), for several combinations of experimental parameters that could be an useful guide in further experimental investigations of the DCE.

We begin by considering a superconducting coplanar waveguide with capacitance and inductance per unit length given, respectively, by $C_0$ and $L_0$,
and terminated at a SQUID (see Ref. \cite{J-R-Johansson-G-Johansson-C-Wilson-F-Nori-PRL-2009-PRA-2010}). Due to the presence of the Josephson junctions in this system,
the electromagnetic field $E(t,x)$ in this coplanar waveguide can conveniently be described by a phase field operator $\phi(t,x) \defeq \int^{t} E(t^{\prime},x) \, dt^{\prime}$, which obeys the massless Klein-Gordon equation $(v^{-2}\;\partial^2_t-\partial^2_x)\phi\left(t,x\right) = 0$, where $v=1/\sqrt{C_0L_0}$ is the speed of light in
the waveguide (unless stated otherwise we assume units in which $\hbar=v=1$). Applying appropriately Kirchoff's laws to the superconducting circuit, it can  be
shown that the BC satisfied by $\phi$, at the origin $x = 0$, is the following one
\cite{wallquist_prb_2006,J-R-Johansson-G-Johansson-C-Wilson-F-Nori-PRL-2009-PRA-2010}
\begin{equation}
\phi\left(t,0\right) = \gamma\left(t\right)
\left[(\partial_x \phi)\left(t,0\right)+\alpha_0(\partial^2_t\phi)\left(t,0\right)\right],
\label{BC}
\end{equation}
where $\alpha_0 = L_0 C$, $C$ is the capacitance of each Josephson junction,
$$
\gamma(t) = - \bar{\Phi}^2_0 \left[ {(2\pi)^2E_{J}(t)L_0} \right]^{-1}=- L_\text{eff}\left(t\right),
$$
where $\bar\Phi_0$ is the magnetic fundamental quantum flux, $E_{J}\left(t\right)$ is the effective Josephson energy
(which depends on the magnetic flux), and $L_\text{eff}\left(t\right)$ is an effective length that modulates the change in time
of the distance between an effective mirror and the SQUID \cite{J-R-Johansson-G-Johansson-C-Wilson-F-Nori-PRL-2009-PRA-2010}.
With the value chosen for the parameter $\alpha_0$ in their experiment, Wilson and collaborators \cite{J-R-Johansson-G-Johansson-C-Wilson-F-Nori-PRL-2009-PRA-2010} neglected in their theoretical calculations the influence of this parameter. However, as we shall see in a moment, slight modifications in the value of $\alpha_0$ lead to quite interesting results. Hence, we adopt the complete BC given by Eq. (\ref{BC}) and analyse its ultimate consequences.
We shall refer to Eq. (\ref{BC}) as the generalized time-dependent Robin BC since it generalizes the time-dependent Robin BC investigated in \cite{Silva-Farina-PRD-2011}. See also \cite{Fosco-et-al-PRD-2013}.

Considering the following general expression for the Josephson energy 
$$E_{J}\left(t\right) =E_{J}^0 \left[1 + \epsilon f(t)\right],$$
with $0 < \epsilon < 1$ and $|f(t)|<1$, we can write the time-dependent Robin parameter as
\begin{equation}
\gamma\left(t\right) \approx \gamma_0 \Bigl[ 1- {\epsilon f\!\left(t\right)}\Bigr],
\label{gamma-of-t}
\end{equation}
where we defined
\begin{equation}
\gamma_0 = - \bar{\Phi}^2_0 \left[ {(2\pi)^2 E_{J}^0 L_0} \right]^{-1}.
\label{gamma-of-zero}
\end{equation}
We shall assume that
$f(t)$ vanishes in both the remote past and distant future.
Using the Ford-Vilenkin perturbative approach \cite{Ford-Vilenkin-PRD-1982}, we express
the field solution as
\begin{equation}
{\phi}\left( t,x\right) \approx
{\phi}_{0}\left( t,x\right) + \epsilon{\phi}_{1}\left( t,x\right),
\label{FV}
\end{equation}
where $\phi_{0}\left(t,x\right)$ is the unperturbed field and
$\epsilon\phi_{1}\left(t,x\right)$ represents the first-order correction due to the time-dependence of $\gamma$,
with $\phi_{0}$ and $\phi_1$ both satisfying the massless Klein-Gordon equation.
The unperturbed field $\phi_{0}$ satisfies the following time-independent generalized Robin BC \cite{Fosco-et-al-PRD-2013} at the origin
\begin{equation}
\phi_0\left(t,0\right)=
\gamma_0 \left[(\partial_x \phi_0)\left(t,0\right)+\alpha_0(\partial^2_t\phi_0)\left(t,0\right)\right],
\label{Time-independent-BC}
\end{equation}
while the BC satisfied by the first order perturbative contribution $\phi_1(t,x)$ can be easily obtained
by substituting (\ref{FV}) in (\ref{BC}) and making use of (\ref{Time-independent-BC}).

Since we are interested in computing the conversion of vacuum fluctuations into real field excitations caused by the time dependence of $\gamma(t)$,
we consider as the initial state of the system (remote past) the vacuum state $\left.|0_{\text{in}}\right\rangle$.

After obtaining the Bogoliubov transformations which relate the creation and annihilation operators in the remote past (``in" operators)
with those in the distant future (``out" operators), and recalling that in the Heisenberg picture the states do not
evolve with time,  the number of created particles between $\omega$ and $\omega+d\omega$ per unit frequency, $N\left(\omega,\gamma_0,\alpha_0\right)$,
can be straightforwardly obtained:
\begin{eqnarray}
N\left(\omega,\gamma_0,\alpha_0\right)  &\defeq& \left\langle 0_{\text{in}}|\right. a^{\dag }_{\text{out}}(\omega)\; a_{\text{out}}(\omega )\left.|0_{\text{in}}\right\rangle
\cr\cr
&=&\frac{2\epsilon^2\gamma_0^2}{\pi}\frac{\omega}{\left[\left(1+\gamma_0\alpha_0\omega^2\right)^2+\omega^2\gamma_0^2\right]}
\int_{-\infty}^{\infty} \!\! \frac{d\omega^{\prime}}{2\pi}
\frac{\Theta\left(-\omega^\prime\right)\left\vert{\omega^\prime}\right\vert\left|{F}(\omega-\omega^{\prime})\right|^2}
{(1+\gamma_0\alpha_0{\omega^\prime}^2)^2+{\omega^\prime}^2\gamma_0^2},
\label{spectral-distribution-general-form}
\end{eqnarray}
where $F(\omega)$ is the Fourier transform of $f(t)$ and $\Theta(\omega)$ is the Heaviside step function. Eq. (\ref{spectral-distribution-general-form}) gives
the perturbative solution for the spectral distribution $N\left(\omega,\,\gamma_0,\,\alpha_0\right)$
up to order $\epsilon$, and for an arbitrary time-dependence of the Robin parameter $\gamma(t) = \gamma_0 \left[ 1 - \epsilon f(t)\right]$ satisfying the
conditions $\vert f(t)\vert < 1$ and $f(t\rightarrow \pm \infty) = 0$. Notice that for $\alpha_0 = 0$,
our result is in perfect agreement with that obtained in \cite{Silva-Farina-PRD-2011}.
The previous work reproduces the results found in \cite{J-R-Johansson-G-Johansson-C-Wilson-F-Nori-PRL-2009-PRA-2010} for 
$\gamma(t) = - \bar{\Phi}^2_0 \left[ {(2\pi)^2E_{J}(t)L_0} \right]^{-1}$.

Hereafter we consider, for practical purposes, a particular but standard oscillatory time variation for the Robin parameter, namely
$
f\!\left(t\right) = \cos \left( \omega_0 t \right) e^{-|{t}|/ \tau},
$
with $\omega_0\tau\gg1$ \cite{J-R-Johansson-G-Johansson-C-Wilson-F-Nori-PRL-2009-PRA-2010,Silva-Farina-PRD-2011}, where $\omega_0$ is the characteristic frequency and $\tau$ is the effective time interval in which the oscillations occur. In an experimental context, this choice of $f(t)$ corresponds to a harmonic change of the magnetic flux through the SQUID.
As a consequence, the Fourier transform $F(\omega)$ is a function  with two sharped peaks around $\omega = \pm \omega_0$ which can be approximated by Dirac delta functions such that
$\left|{F}(\omega)\right|^2 \approx \left({\pi}/{2}\right) \tau \left[ \delta(\omega-\omega_0)+\delta(\omega+\omega_0) \right]$.
Substituting this expression into (\ref{spectral-distribution-general-form}) and recovering the speed of light in the waveguide $v$,  the ratio $\mathcal{N} \defeq N/\tau$ is found to be
%
\begin{equation}
{\mathcal{N}\left(\omega,\gamma_0,\alpha_0\right)} =
\frac{\epsilon^2\gamma_0^2}{2\pi v^2}
\frac{\omega\left(\omega_0-\omega \right)}{\left(1+\gamma_0\alpha_0\omega^2\right)^2+
\frac{\gamma_0^2\omega^2}{v^2}}
\frac{\Theta \left(\omega_0-\omega \right)}
{\left[1+\gamma_0\alpha_0\left(\omega_0-\omega \right)^2 \right]^2
+\frac{\gamma_0^2\left(\omega_0-\omega \right)^2}{v^2}}.
\label{spectral-distribution-correct-dimension}
\end{equation}
This is our main result and, as we will show, this formula reveals remarkable features that may be useful in future experiments to observe the DCE in the context of circuit QED, the most surprising one being the possibility of breakdown of the parabolic spectral distribution for appropriate values of $\alpha_0$ and $\gamma_0$.
However, let us first make a few comments about the above spectral distribution.
Note that for any values of $\gamma_0$ and $\alpha_0$,
 $\mathcal{N}\left(\omega,\gamma_0,\alpha_0\right)$ is invariant under the transformation $\omega \rightarrow \omega_0 - \omega $,
which means that it is symmetric with respect to $\omega = \omega_0/2$.
Moreover, because of the Heaviside step function, no particles are created with frequencies $\omega >\omega_0$. Finally, for $\alpha_0 = 0$,  Eq. (\ref{spectral-distribution-correct-dimension})
coincides with the previously known result obtained in \cite{Silva-Farina-PRD-2011}, as expected.
Integrating the spectral distribution for all frequencies, we obtain the particle creation rate $\mathcal{R}$, namely,
\begin{equation}
\mathcal{R}\left(\omega_0,\gamma_0,\alpha_0\right) \defeq
\int_{0}^{\infty}\!\!\! d\omega \;\mathcal{N}\left(\omega,\gamma_0,\alpha_0\right).
\label{GR-particle-creation-rate}
\end{equation}

Now, let us investigate how the spectral distribution $\mathcal{N}\left(\omega,\gamma_0,\alpha_0\right)$
 given by Eq. (\ref{spectral-distribution-correct-dimension}) depends on parameters $\alpha_0$ and $\gamma_0$
(keeping $v$ constant).
 Particularly, since we are searching for new signatures of the DCE, we want to determine which values of $\alpha_0$ and $\gamma_0$ may cause
 a substantial change in the spectral distribution and consequently in the particle creation rate. With this purpose in mind, we shall compare the value of $\mathcal{N}\left(\omega,\gamma_0,\alpha_0\right)$ given by Eq.(\ref{spectral-distribution-correct-dimension}),
 for a given frequency (we choose it to be half the driven frequency $\omega_0/2$), with that obtained by taking for $\alpha_0$ and $\gamma_0$ the
 values adopted in \cite{C-Wilson-et-al-Nature-2011}. The values for $\alpha_0$ and $\gamma_0$ used in this experiment, relabeled conveniently as
$\alpha_{0\text{exp}}$ and $\gamma_{0\text{exp}}$, are given by $\alpha_{0\text{exp}}=0.41 \times 10^{-19}\;\text{s}^2\text{/m}$ and $\gamma_{0\text{exp}}=-0.44 \,\times 10^{-3} \,\text{m}$. Hereafter we also consider the following values for the other relevant quantities for the SQUID experiment: $\omega_0=2\pi \times 10.30$ GHz, $\epsilon=0.25 $ and $v = 1.2 \times 10^{8}$ m/s \cite{J-R-Johansson-G-Johansson-C-Wilson-F-Nori-PRL-2009-PRA-2010, C-Wilson-et-al-Nature-2011}. Additionally, it is convenient to parametrize our problem in terms of the dimensionless variables $\xi \defeq \gamma_0/\gamma_{0\text{exp}}$ and $\zeta \defeq \alpha_0/\alpha_{0\text{exp}}$. Hence, from now on, we shall write $\mathcal{N}(\omega,\,\xi,\,\zeta)$ and $\mathcal{R}(\omega_0,\,\xi,\,\zeta)$ instead of
 $\mathcal{N}\left(\omega,\gamma_0,\alpha_0\right)$ and $\mathcal{R}\left(\omega_0,\gamma_0,\alpha_0\right)$. We emphasize that the spectral distribution and the particle creation correspondent to the experimental values are given,
respectively, by $\mathcal{N}\left(\omega,1,1\right)$ and $\mathcal{R}\left(\omega_0,1,1\right)$.

\begin{figure}[!hbt]
\centering
\includegraphics[scale=0.27]{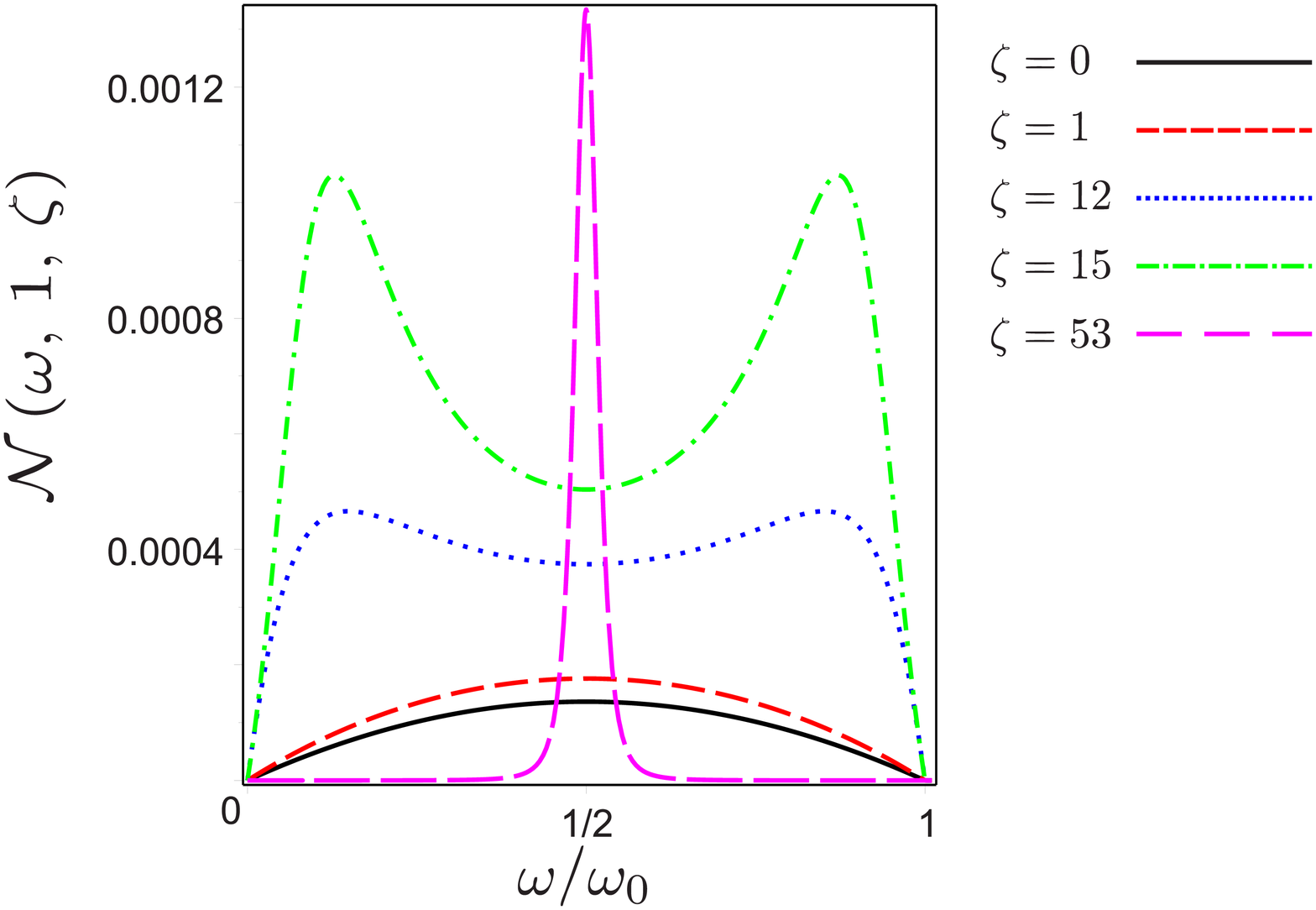}
\caption{(Color online). Spectral distributions for $\xi = 1$ and different values of $\zeta$. Note the
two peak pattern for   $\zeta = 12$  (blue dotted line) and $\zeta = 15$ (green dashed-dotted line).
For $\zeta \approx 53$ (purple large-dashed line) we conveniently multiplied the values of
 ${\mathcal{N}\left(\omega,\,1,\,53\right)}$  by a factor of $2 \times 10^{-3}$. }
\label{espectro-gamma0-1}
\end{figure}

Let us start by studying the role played by $\alpha_0$. Henceforth, we assume a fixed value of $\gamma_0$, say $\gamma_0 = \gamma_{0\text{exp}}$, and vary only the parameter $\alpha_0$.
 In Fig. \ref{espectro-gamma0-1} we plot the spectral distribution for different values of $\alpha_0$, i.e., we plot $\mathcal{N}\left(\omega, \,1,\,\zeta\right)$ versus $\omega/\omega_0$ but keeping $\xi = 1$ while the value of $\zeta$ is increased 
 (it is important to have in mind  that $\xi = 1$ means a negative value for $\gamma_0$, namely, 
 $\gamma_0 = \gamma_{0exp} = -0,44 \times 10^{-3} $m).

%

First, note the symmetry with respect to $\omega/\omega_0 = 1/2$, as expected.
For $\zeta=0$ the spectrum is characterized by the parabole showed by the solid line. In theoretical calculations related to the SQUID experiment \cite{J-R-Johansson-G-Johansson-C-Wilson-F-Nori-PRL-2009-PRA-2010},
$\alpha_0$ was neglected, so that the solid line in Fig. \ref{espectro-gamma0-1} corresponds to the theoretical spectral distribution
predicted in \cite{J-R-Johansson-G-Johansson-C-Wilson-F-Nori-PRL-2009-PRA-2010}.
If we do not neglect $\alpha_0$ (i.e\, we assume $\zeta \neq 0$), but consider its value assumed in the SQUID experiment ($\zeta = 1$),
though the spectrum is still characterized by a parabolic curve as shown by the dashed line,
 its maximum value is $8\%$ greater than the corresponding maximum value with $\zeta = 0$.
 However, as we increase $\zeta$ a quite unexpected feature appears: for $\zeta = 12$ (dotted line) and $\zeta = 15$ (dashed-dotted line), we see a \textit{departure} from the usual parabolic spectrum with the appearance of two peaks, as well as an \textit{enhancement}
of  $\approx 295\%$ ($\zeta = 12$) and $\approx 578\%$ ($\zeta = 15$)
(in comparison with the experimental values obtained with $\xi=1$ and $\zeta=1$)
in the particle creation rate  $\mathcal{R}$ (the area between the horizontal axis and each curve in Fig. \ref{espectro-gamma0-1}).

Increasing further the value of $\zeta$, instead of a two peaked spectral distribution, we see a \textit{narrow sharp peak} centered at $\omega / \omega_0 = 1/2$, as shown in Fig. \ref{espectro-gamma0-1} for $\zeta = 53$ (large dashed line).
This peak is $\approx 10^3$ times larger than the maximum value of other curves displayed in Fig. \ref{espectro-gamma0-1}, and there is an increase in the particle creation rate by a factor of $\approx 3 \times 10^2$ as we shall see in a moment. The appearance of the two peaked curve for $\mathcal{N}(\omega,\,\xi,\,\zeta)$, resembling the humps of a camel, and also the appearance of a sharp peak, provide new signatures of the DCE in superconducting circuits.

\begin{figure}[h]
\centering
\includegraphics[scale=0.26]{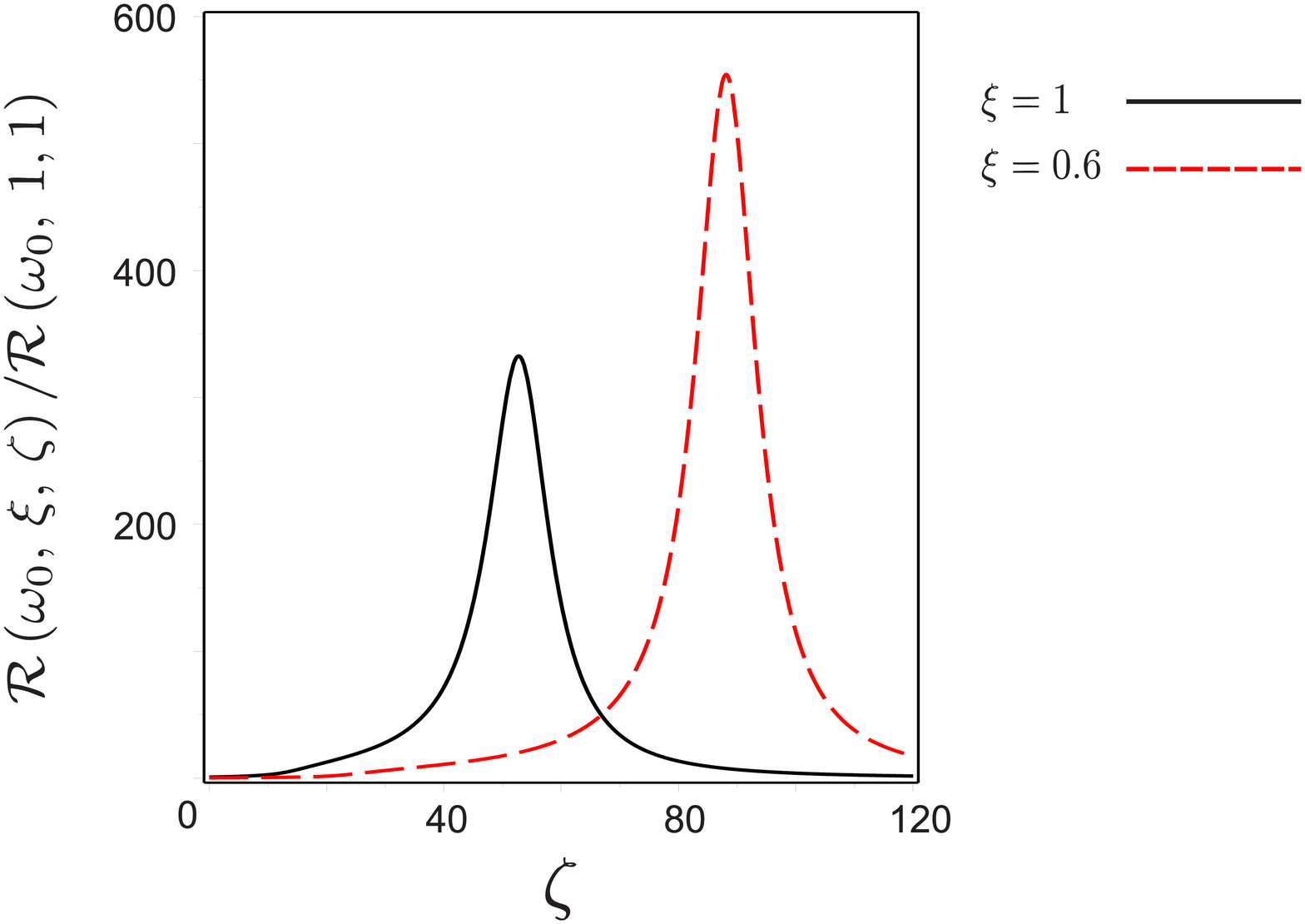}
\caption{(Color online) Particle creation rate $\mathcal{R}\left(\omega_0,\,\xi,\,\zeta\right)$, normalized by $\mathcal{R}\left(\omega_0,\,1,1\right)$, as a function of $\zeta$ with two fixed values for $\xi$. For $\xi = 1$ the the maximum occurs for $\zeta \approx 53$ as anticipated from Fig. (\ref{espectro-gamma0-1}). For $\xi = 0.6$ the maximum occurs for $\zeta \approx 88$ and implies in a dramatic increase in the particle creation rate by a factor of $\approx 5\times 10^2$.}
\label{ParticleCreationRate}
\end{figure}

Let us now quantify how the particle creation rate $\mathcal{R}\left(\omega_0,\xi,\zeta\right)$ depends on
parameter $\zeta$, with fixed $\xi$, in order to investigate real possibilities of substantial intensification of this rate.
This could potentially be relevant in conceiving future experiments to observe the DCE. In Fig. \ref{ParticleCreationRate} we show the ratio $\mathcal{R}\left(\omega_0,\,\xi,\,\zeta\right)/\mathcal{R}\left(\omega_0,\,1,1\right)$ as a function of $\zeta$.
%
%
It reveals that as  $\zeta$ is increased from zero, the ratio
${\mathcal{R}\left(\omega_0,\,\xi,\,\zeta\right)}/{\mathcal{R}\left(\omega_0,\,1,1\right)}$
increases monotonically until it reaches a maximum value and from this point on it decreases monotonically.
Observe that for each value of $\xi$ there is a value of $\zeta$ for which the total particle creation rate reaches a maximum.
This maximum value is $\approx 3\times 10^2$ (for $\xi=1$) and $\approx 5\times 10^2$ (for $\xi=0.6$) times greater than ${\mathcal{R}\left(\omega_0,\,1,1\right)}$.

\begin{figure}[!hbt]
\centering
\includegraphics[scale=0.25]{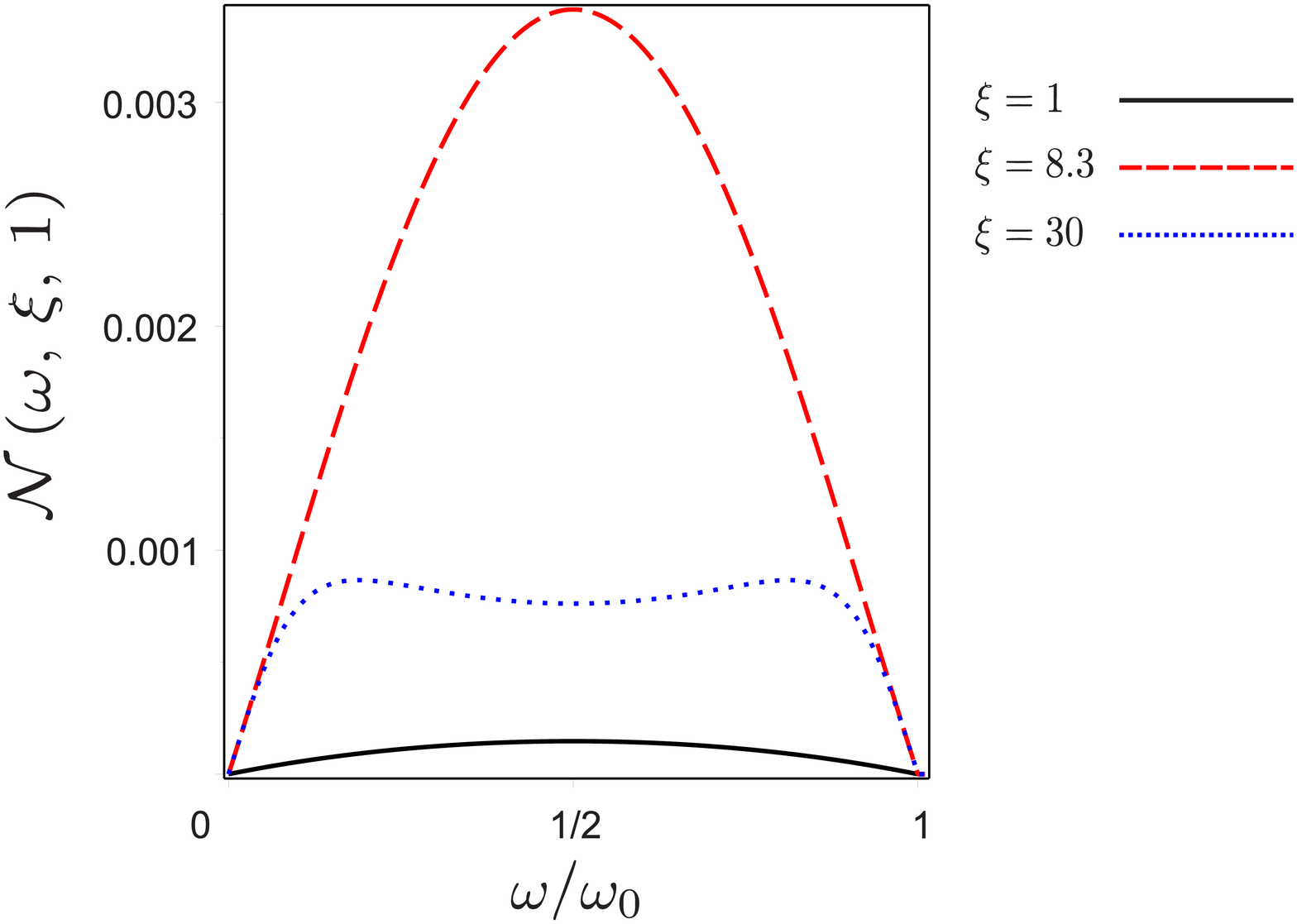}
\caption{(Color online). ${\mathcal{N}\left(\omega,\,\xi,\,1\right)}$ versus $\omega/\omega_0$ for
different values of $\xi$:  solid line curve for $\xi = 1$;
dashed line curve for $\xi = 8.3$ and dotted line curve for $\xi = 30$.}
\label{espectro-alpha0-1}
\end{figure}

We can also make a similar analysis of the spectral distribution as a function of $\xi$, but now with a fixed value for $\zeta$, chosen
conveniently as $\zeta = 1$.
As an example, in Fig. \ref{espectro-alpha0-1} we plot ${\mathcal{N}\left(\omega,\,\xi,\,1\right)}$
versus $\omega/\omega_0$ for different values of $\xi$.
For $\xi=1$, we have the curve corresponding to the experimental values.
Enhancing $\xi$, we have an enhancement of the area between the horizontal axis and the curve
of the spectral distribution
(in other words, an enhancement of $\mathcal{R}$).
The largest $\mathcal{R}$ is obtained with $\xi\approx 8.3$. Then,
for larger values of $\xi$ the area diminishes, with the appearance of
double peaked curves (for instance, for $\xi = 30$).
%

\begin{figure}[!hbt]
\centering
\includegraphics[scale=0.40]{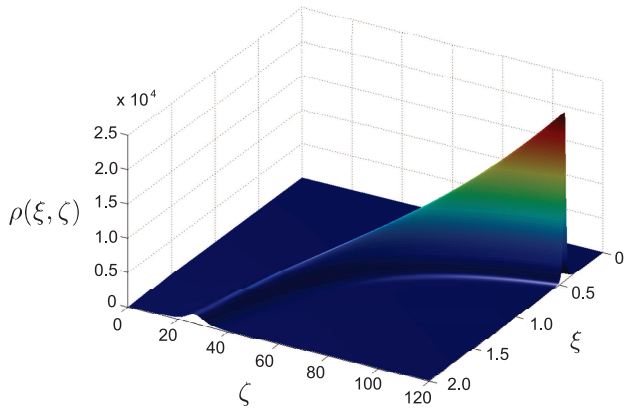}
\caption{(Color online). Ratio $\rho(\xi,\zeta)$ as a function of $\xi$ and $\zeta$. For $\xi \approx 0.6$ and $\zeta\approx 88$ we find that $\rho(\xi,\zeta) \approx 1.3 \times 10^{4}$.}
\label{cordilheira}
\end{figure}

From our analysis of the Fig. \ref{espectro-gamma0-1}
and \ref{espectro-alpha0-1}, we see that the maximum value of
$\mathcal{R}$ (the largest area between the horizontal axis and a given curve) is associated with
the situation where $\mathcal{N}$ at $\omega_0/2$ has its maximum value.
Then, to identify regions in the configuration space $(\xi,\zeta)$ where
$\mathcal{R}$ is large, we investigate the behavior of the creation of particles with frequency around $\omega = \omega_0 / 2$ (half the driven frequency).
Fig. \ref{cordilheira} shows the behavior of $\rho(\xi,\zeta)$, defined as,
\begin{equation}
\rho(\xi,\zeta) \defeq \frac{{\mathcal{N}\left(\omega_0/2,\,\xi,\,\zeta\right)}}
{{\mathcal{N}\left(\omega_0/2,\,1,\,1\right)}},
\end{equation}
as a function of $\xi$ and $\zeta$,
highlighting the region where particle creation is more relevant.
For instance, for $\xi \approx 0.6$ and $\zeta\approx 88$, an enhancement by a factor of $10^4$ in the number of created particles with  $\omega = \omega_0/2$
can be seen in Fig. \ref{cordilheira}.

%

\begin{figure}[!hbt]
\centering
\includegraphics[scale=0.24]{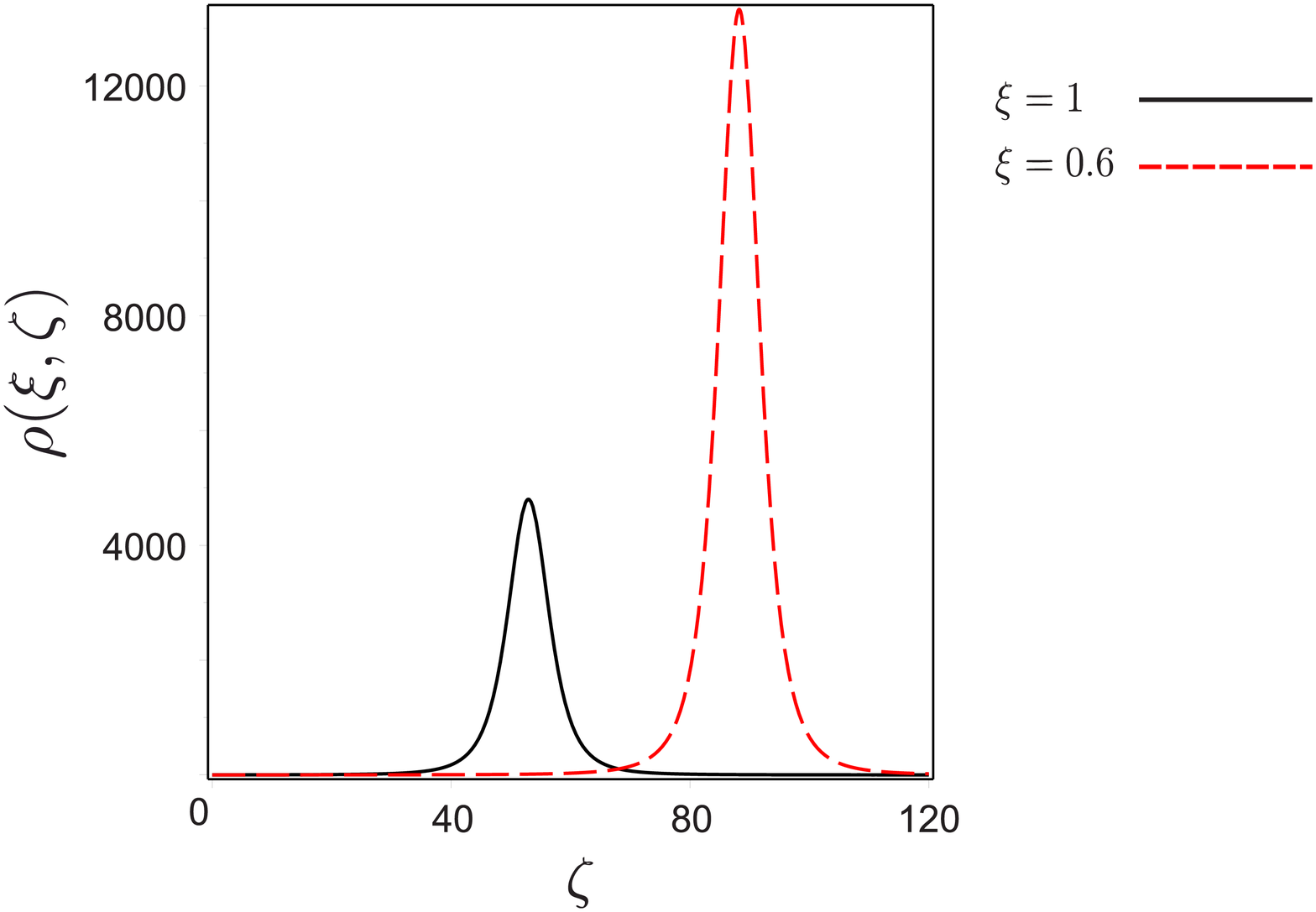}
\caption{(Color online) The ratios $\rho(1,\zeta)$ (black solid line) and $\rho(0.6,\zeta)$ (red dashed line)  as functions of $\zeta$.}
\label{corte-xi-const}
\end{figure}

An inspection in this figure shows that for a fixed $\xi$, $\rho$ exhibits  a maximum value for a given $\zeta$.
This is  shown more explicitly in Fig. \ref{corte-xi-const}, where we plot $\rho(\xi,\zeta)$
as a function of $\zeta$ for two fixed values of $\xi$.
The solid line corresponds to $\xi = 1$ and, for this case, the maximum value of $\rho$ ($\approx 5\times10^{3}$) occurs for $\zeta \approx 53$.
The dashed line corresponds to $\xi = 0.6$ and has a maximum value (about $\approx 1.3\times10^{4}$)
for $\zeta \approx 88$. Though these figures give information only about particle creation with $\omega = \omega_0/2$, it can be shown that
also the particle creation rate $\mathcal{R}$ is enhanced for these values of $\xi$ and $\zeta$.

In conclusion, we have shown that appropriate adjustments of the parameters used in the SQUID experiment surprisingly reveals at least three remarkable predictions, namely:
 unexpected non-parabolic spectral distributions; an enhancement in the created particles with $\omega=\omega_0/2$ by a factor that can reach up to $5\times10^3$ for $\xi=1$ and $\zeta \approx 53$ (or, for instance, $1.3 \times 10^{4}$ for $\xi=0.6$ and $\zeta \approx 88$) and an increase in the particle creation rate by a factor greater than $3\times10^2$ with respect to ${\mathcal{R}\left(\omega_0,\,1,1\right)}$.
We think these theoretical predictions can be viewed as new evidences for the DCE in superconducting circuits and can be helpful for conceiving future experiments.  The plot shown in Fig. \ref{cordilheira} acts as a guide for the  experimentalists who will be able to look for the highest peak that can be achieved with feasible values for the parameters $\gamma_0$ and $\alpha_0$.  We hope that the consideration of different values for these parameters  may be of some help in the identification of dynamical Casimir photons in future experiments.

\section{Acknowledgments}
\label{ack}

We are grateful to C. M. Wilson for valuable discussions at the  Pan-American Advanced Study Institute during the conference ``{\it Frontiers of Casimir Physics}\,".
The authors are indebted to P. A. Maia Neto and F. D. Mazzitelli for enlightening discussions. We also acknowledge G. S. Paraoanu for bringing Ref. \cite{Lahteenmaki-et-al-PNAS-2013} to our attention and Clerisson M. do Nascimento for generating Fig. (\ref{cordilheira}). A.L.C.R. thanks Universidade Federal do Par\'a for the hospitality. This work was
partially supported by the brazilian agencies CNPq, CAPES and FAPERJ.
%
%


\begin{thebibliography}{99}
%
\bibitem{Moore-1970} G. T. Moore, {J. Math. Phys.} {\bf 11}, 2679 (1970).

\bibitem{Dewitt-PhysRep-1975} B.S. DeWitt, {Phys. Rep.} {\bf 19}, 295 (1975).


\bibitem{Fulling-Davies-PRS-1976-I-Fulling-Davies-PRS-1977-I}
S.A. Fulling, and P.C.W. Davies, {Proc. R. Soc. London} {\bf A 348}, 393 (1976); P.C.W. Davies, and S.A. Fulling, {Proc. R. Soc. London} {\bf A 354}, 59 (1977).

%
\bibitem{Kim-Brownell-Onofrio-2006} W. J. Kim, J. H. Brownell and R. Onofrio , {Phys. Rev. Lett.} {\bf 96}, 200402 (2006).
%
\bibitem{Yablonovitch-1989}  E. Yablonovitch, {Phys. Rev. Lett.} {\bf 62}, 1742 (1989).
%
\bibitem{experimentos-Braggio} C. Braggio {\it et al}, {Europhys. Lett} {\bf 70}, 754 (2005); A. Agnesi {\it et al}, {J. Phys.} A {\bf 41}, 164024 (2008); A. Agnesi {\it et al}, {J. Phys: Conf. Series} {\bf 161}, 012028 (2009).
%
\bibitem{J-R-Johansson-G-Johansson-C-Wilson-F-Nori-PRL-2009-PRA-2010} J. R. Jo\-han\-sson, G. Jo\-han\-sson, C. M. Wilson, and F. Nori, {Phys. Rev. Lett.}  {\bf 103}, 147003 (2009); {Phys. Rev. A}  {\bf 82}, 052509 (2010).
%
\bibitem{Dezael-Lambrecht-EPL-2010} F. X. Dezael, and A. Lambrecht, {Eur. Phys. Lett.} {\bf 89}, 14001 (2010).
%
\bibitem{Kawakubo-Yamamoto-PRA-2011} T. Kawakubo, and K. Yamamoto, {Phys. Rev.} A, {\bf 83}, 013819 (2011).
%
\bibitem{Faccio-Carusotto-EPL-2011}  D. Faccio and I. Carusotto, {Eur. Phys. Lett} {\bf 96}, 24006 (2011).
%
\bibitem{C-Wilson-et-al-Nature-2011} C. M. Wilson {\it et al}, {Nature (London)} {\bf 479}, 376 (2011).
%
\bibitem{Lahteenmaki-et-al-PNAS-2013} P. L{\"a}hteenm{\"a}ki, G. S. Paraoanu, J. Hassel, and P. J. Hakonen, {Proc. Natl. Acad. Sci.} {\bf 110}, 4234 (2013).
%
\bibitem{Lambrecht-Jaekel-Reynaud-PRL-1996} A. Lambrecht, M. T. Jaekel, and S. Reynaud, {Phys. Rev. Lett.}  {\bf 77}, 615 (1996).
%

\bibitem{Lambrecht-et-al-EPJD-1998} A. Lambrecht M. T. Jaekel, and S. Reynaud, {Eur. Phys. J. D}  {\bf 3}, 95 (1998).
%
\bibitem{wallquist_prb_2006} M. Wallquist, V. S. Shumeiko and G. Wendin, {Phys. Rev. B} {\bf 74}, 224506 (2006).
%
\bibitem{Ford-Vilenkin-PRD-1982} L. H. Ford, and A. Vilenkin, {Phys. Rev. D} {\bf 25}, 2569 (1982).
%
\bibitem{Silva-Farina-PRD-2011} H. O. Silva, and C. Farina, {Phys. Rev. D} {\bf 84}, 045003 (2011).
%
\bibitem{Fosco-et-al-PRD-2013} C. D. Fosco, F. C. Lombardo, and F. D. Mazzitelli,  {Phys. Rev. D} {\bf 87}, 105008 (2013).
%
%
\end{thebibliography}
\end{document}